\documentclass[useAMS,usenatbib]{mn2e}
\usepackage{graphicx,natbib,times,mathptm}

\def\Halpha{{\rm H}\alpha}

\bibpunct{(}{)}{;}{a}{}{,}

\begin{document}

\bibliographystyle{mn2e}

\title[Structure of the Outer Galactic Disc]{The Structure of the Outer Galactic Disc as revealed by IPHAS early A Stars}

\author[Sale et al.]{S. E. Sale$^1$\thanks{E-mail: s.sale06@imperial.ac.uk}, J. E. Drew$^{2}$, C. Knigge$^3$, A. A. Zijlstra$^{4}$, M. J. Irwin$^5$, R. A. H. Morris$^{6}$, 
\newauthor S. Phillipps$^{6}$, J. J. Drake$^{7}$, R. Greimel$^{8}$, Y. C. Unruh$^{1}$, P. J. Groot$^{9}$, A. Mampaso$^{10}$, 
\newauthor N. A. Walton$^{5}$ \\
$^1$ Astrophysics Group, Imperial College London, Blackett Laboratory, Prince Consort Road, London SW7~2AZ, U.K.\\
$^2$ Centre for Astrophysics Research, STRI, University of Hertfordshire, College Lane Campus, Hatfield, AL10~9AB, U.K.\\
$^3$ School of Physics \& Astronomy, University of Southampton, Southampton, SO17 1BJ, U.K. \\
$^4$ Jodrell Bank Center for Astrophysics, Alan Turing Building, The University of Manchester, Oxford Street, Manchester, M13~9PL, U.K. \\
$^5$ Institute of Astronomy, Madingley Road, Cambridge CB3~0HA, U.K.\\
$^6$ Astrophysics Group, Department of Physics, Bristol University, Tyndall Avenue, Bristol, BS8 1TL, U.K.\\
$^{7}$ Harvard-Smithsonian Center for Astrophysics, 60 Garden Street, Cambridge, MA 02138, USA\\
$^8$ Institut f\"ur Physik, Karl-Franzens Universit\"at Graz, Universit\"atsplatz~5, 8010 Graz, Austria\\
$^9$ Department of Astrophysics/IMAPP, Radboud University Nijmegen, PO Box 9010, 6500 GL, Nijmegen, the Netherlands\\
$^{10}$ Instituto de Astrof\'{\i}sica de Canarias, Via L\'actea s/n, E38200 La Laguna, Santa Cruz de Tenerife, Spain \\
}

\date{Received .........., Accepted...........}

\maketitle

\begin{abstract}
This study is an investigation of the stellar density profile of the Galactic disc in the Anticentre direction.  We select over 40,000 early A stars from IPHAS photometry in the Galactic longitude range $160^{\circ} < l < 200^{\circ}$ close to the equatorial plane ($-1^{\circ} < b < +1^{\circ}$). We then compare their observed reddening-corrected apparent magnitude distribution with simulated photometry obtained from parameterised models in order to set constraints on the Anticentre stellar density profile. By selecting A stars, we are appraising the properties of a population only $\sim 100$~Myrs old.  We find the stellar density profile of young stars is well fit to an exponential with length scale of $(3020 \pm 120_{statistical} \pm 180_{systematic})$~pc, which is comparable to that obtained in earlier studies, out to a Galactocentric radius of $R_T = (13.0 \pm 0.5_{statistical} \pm 0.6_{systematic})$~kpc.  At larger radii the rate of decline appears to increase with the scale length dropping to $(1200 \pm 300_{statistical} \pm 70_{systematic})$~pc. This result amounts to a refinement of the conclusions reached in previous studies that the stellar density profile is abruptly truncated. The IPHAS A star data are not compatible with models that propose a sudden change in metallicity at $R_G = 10$~kpc.

\end{abstract}

\section{Introduction}

Our position within the Milky Way Galaxy makes studying its structure and evolution both uniquely challenging and rewarding. Its constituent components can be observed in unique detail. However, disentangling its structure is complicated by difficulties in obtaining accurate distances to objects and frequently heavy extinction. Studying the Galactic thin disc is particularly difficult; within the thin disc extinction is at its strongest and its constituent stellar population is a smooth mix of stars formed over a wide range of ages, in vastly differing conditions and locations, and which migrate across the Galaxy.

Such difficulties help to explain the vast range of results which have been obtained when determining parameters such as the scale length of the stellar density in the thin disc. Early estimates tended to be relatively long, for example: \cite{deVaucouleurs_Pence.1978} obtained 3800~pc, \cite{vanderKruit_only.1986} 5500~pc, \cite{Habing_only.1988} 4500~pc, \cite{Lewis_Freeman.1989} 4400~pc and \cite*{Kent_Dame.1991} 3.0~kpc. \cite*{Robin_Creze.1992} compared their Galactic model to CCD photometry of a 29 ~arcmin$^2$ area in SA23 and found a considerably shorter scale length of 2500~pc. \cite{Ruphy_Robin.1996} adopted a similar approach, but this time using photometry from the DENIS survey, yielding a scale length of 2250~pc, whilst, \cite{Robin_Reyle.2003} obtained a scale length of 2530~pc from DENIS data. \cite{Freudenreich_only.1996} derived a length of 2600~pc from DIBRE observations. \cite*{Buser_Rong.1998} found 4000~pc and \cite{Siegel_Majewski.2002} found 2500--3125~pc, once a correction for binarity has been applied. Most recently \cite{Juric_others.2008} measured the thin disc scale length, using high Galactic latitude photometry from SDSS, finding a scale length of 2600~pc. Many of the previous studies have not used observations at low latitude, reducing the sensitivity of their measurements and preventing them from studying the disc at larger Galactocentric radii.

It is important to note that the scale length of the thin disc in spiral galaxies is expected to change: it may increase with time due to so called inside--out galaxy formation. This process involves the galaxy accreting gas in its outer disc at later times, from which it forms new stars, whilst simultaneously exhausting gas supplies in its core. Inside--out formation is discussed in \cite{Larson_only.1976}, \cite{Matteucci_Francois.1989}, \cite*{Chiappini_Matteucci.1997} and many others. Within the local group, it has been suggested that inside--out formation would account for observed metallicity and mean stellar age gradients in M33 \citep{Magrini_Corbelli.2007, Verley_Corbelli.2009, Williams_Dalcanton.2009}. In our own Galaxy, several authors \citep[including:][]{Cescutti_Matteucci.2007, Colavitti_Cescutti.2009, Magrini_Sestito.2009} have demonstrated that chemodynamical models of the formation and evolution of the Galaxy, which incorporate inside-out formation, produce results which are consistent with some observations of radial abundance gradients. 

The effect of studying a Galaxy with a thin disc scale length which varies with time is that any measurement of the scale length from stars will be dependent on the types of stars used. If stars could be easily grouped by age, one would expect the younger groups to show a longer scale length than the older ones if the Galaxy was undergoing inside--out formation. Unfortunately, in practice it is not possible to determine accurate ages for most stars and so any samples selected on the basis of spectral type for example will contain a superposition of ages and thus the measured scale length will be dependent on the range of ages covered by the sample.

\cite{vanderKruit_only.1979} first observed that the exponential discs of some spiral galaxies are radially truncated, Indeed it has subsequently been observed that truncation is present in the light profiles of the majority of spiral galaxies \citep*{Pohlen_L"utticke.2001, Kregel_vanderKruit.2002}. With local galaxies it is possible to resolve individual stars and so study the radial distribution of a given range of stars. M33 \citep{Ferguson_Irwin.2007, Williams_Dalcanton.2009} shows truncation in the radial distribution of predominantly RGB stars. Other local galaxies, such as NGC 300 \cite*{Bland-Hawthorn_Vlajic.2005}, appear not to be truncated. In this context, the truncation of the Galactic disc, as shown by \cite{Robin_Creze.1992} and \cite{Ruphy_Robin.1996} is not surprising. The key difference between their results and studies of other galaxies, is that they find that their observations are well described by a model Galactic disc which is sharply truncated, with stellar density dropping to zero beyond some Galactocentric radius, rather than a model where the truncation is less sharp. It is worth noting, that although a model with such a sharp truncation is a good fit to their observations and a better fit than no truncation, it is possible that a model with a less sharp truncation would also fit their observations. \cite{Ruphy_Robin.1996} find that the truncation occurs at a radius of $15 \pm 2$~kpc, whilst \cite{Robin_Creze.1992} find that a model with truncation at a radius of 14~kpc best describes their observations.

The cause of radial truncation in galactic discs is poorly understood \citep[e.g.][]{Elmegreen_Hunter.2006}. There have been several mechanisms proposed to explain it. \cite{Kennicutt_only.1989} suggested that star formation in a gaseous disc may be truncated when the column density of the gas drops bellow the threshold for gravitational instabilities to form, based on the sharp truncation of $\Halpha$ emission observed in many discs. \cite{Boissier_Prantzos.2003} were able to show that truncation of the form proposed by \cite{Kennicutt_only.1989} was consistent with their observations of the Galaxy. \cite{Naab_Ostriker.2006} suggested that truncation due to this mechanism would occur at a Galactocentric radius of $\sim 12 kpc$ in the Galaxy. \cite{Elmegreen_Parravano.1994} and \cite{Schaye_only.2004} suggest that the transition of the ISM from being dominantly in the cold phase to being dominantly in the warm phase is the more correct physical factor suppressing star formation. A very different view has been taken by \cite{vanderKruit_only.1987}, who suggested that the gaseous disc of a galaxy may be truncated during the galaxy's initial formation. Further mechanisms for truncating the disc involve magnetic fields \citep*{Battaner_Florido.2002} and the outer Linblad resonance \citep*{Erwin_Pohlen.2008}. 

\cite{Rovskar_Debattista.2008} demonstrate how the sharpness of disc truncation may be reduced by the redistribution of stars as a result of secular processes in the disc. \cite*{Azzollini_Trujillo.2008} show, using $B$-band photometry, that the truncation radii of galaxies increase with age, whilst the surface brightness at the break decreases. \cite{Pflamm-Altenburg_Kroupa.2008} suggest it is possible for a galaxy to exhibit sharp disc truncation in $\Halpha$ emission without this suppression of star formation. 

It is important to be aware of what is viewed as truncated in each study. Observations of the Galaxy and other local galaxies are able to study the truncation in the numbers of certain types of stars. In more distant galaxies the ability to resolve individual stars is lost. As a result, it is the light profile of more distant galaxies that is frequently examined for truncations. Both \cite{Martinez-Serrano_Serna.2009} and \cite{Sanchez-Blazquez_Courty.2009} demonstrate models of galaxy formation which can produce galaxies with truncated light profiles and untruncated stellar mass profiles. The difference in the profiles is the result of the mean age of stars varying with radius, beyond the observed truncation radius in the light profiles the mean age of stars increases resulting in a smaller number of young and bright early-type stars. These models are in line with the observations of \cite*{Bakos_Trujillo.2008} who observe consistent $(g'-r')$ profiles. Therefore, it is necessary to study truncation in the discs of galaxies with resolved stellar populations so that the effects of changing stellar age and density can properly be understood. Of the galaxies in which we can resolve their stellar populations it is our own which can be studied in the greatest detail.

The radial dependence of galactic metallicity is also unclear. Metallicity may drop with a constant gradient \citep[e.g.][]{Friel_Janes.2002, Bragaglia_Tosi.2006, Luck_Kovtyukh.2006}, or the gradient may flatten at large Galactocentric radii \citep{Carraro_Geisler.2007, Bragaglia_Sestito.2008, Lemasle_others.2008}. The latter option would agree with the chemical evolution models of \cite{Chiappini_Matteucci.1997} and \cite{Cescutti_Matteucci.2007}. These studies assume the inside-out formation of the galactic disc proceeded via two episodes of significant material infall, the first being associated with the formation of the Galaxy. \cite{Maciel_Costa.2009} estimated the time evolution of the Galactic metallicity gradient, based on central stars of planetary nebulae with a range of ages. They conclude that the gradient appears to have been steeper in the past, a finding which would agree with inside-out formation. An alternative scheme is that the metallicity obeys a step function, dropping suddenly at $R_G \approx 10$~kpc \citep*{Corder_Twarog.2001, Yong_Carney.2005}, this could arise as the result of a significant accretion event. The tracers used to study metallicity will affect the results obtained: the use of relatively old objects \citep[e.g. old open clusters in][]{Bragaglia_Tosi.2006} compared to younger ones \citep[e.g. OB stars in][]{Daflon_Cunha.2004} will probe the metallicity in the ISM at an earlier age. Thus, as the metallicity of the ISM will have evolved over time, the results obtained will be different.  Even where the choice of scheme agrees and the same tracers are used there is still significant disagreement on the size of any gradient \cite[and references therein]{Lemasle_others.2008}. 

Studies of other spiral galaxies have revealed that they too exhibit metallicity gradients. M31 appears to show a metallicity gradient \citep{Worthey_Espana.2005}, though analysing it is complicated by the inclination of M31 and difficulties in separating contributions from its disc and bulge components. Due to its more favourable inclination M33 is frequently considered to be better suited for studying abundance gradients and has long been known to posses them \citep[e.g.][]{Searle_only.1971}. More recently these have been studied in greater detail, using a range of tracers, for example: planetary nebulae \citep[e.g.][]{Magrini_Perinotto.2004}; red giant branch stars \citep{Barker_Sarajedini.2007}; HII regions \citep{Magrini_Vilchez.2007}. As with our Galaxy, the results obtained for M33 can diverge, even if the same tracer has been employed \citep[][and references therein]{Magrini_Vilchez.2007}. \cite{Magrini_Stanghellini.2009} determine that the metallicity gradients in M33 do not appear to have changed significantly with time. Outside of the Local Group, metallicity gradients have been recorded in other spiral galaxies, such as NGC 300 \citep{Vlajic_Bland-Hawthorn.2009}. The possible flattening of the metallicity gradient at large radii does not appear to be a unique feature of our Galaxy: similar behaviour has been observed in M31, M83 and NGC 300 \citep{Worthey_Espana.2005, Bresolin_Ryan-Weber.2009, Vlajic_Bland-Hawthorn.2009}.

We investigate the outer disc by comparing the photometry of A stars derived from IPHAS \citep{Drew_others.2005}, with simulations exploiting aspects of the Besan\c{c}on formulation of Galactic models \citep{Robin_Reyle.2003}. In section \ref{observations} we introduce the CCD photometry we have used and detail how it has been prepared. Section \ref{method} explains why and how observed and simulated photometry are compared. The results obtained are presented in section \ref{results} along with discussion and analysis of other relevant parameters.

\section{Observations}\label{observations}

The INT/WFC Photometric $\Halpha$ Survey of the northern Galactic Plane \citep[IPHAS; ][]{Drew_others.2005} is the first comprehensive digital survey of the disc of the Galaxy ($|b|\leq5^{\circ}$), north of the celestial equator. Imaging is performed in the $r'$, $i'$ and $\Halpha$ bands down to $r'\sim20$ ($10\sigma$). At the time of writing all fields in the survey area have been observed at least once, with a minority of these in need of replacement by data obtained in better observing conditions.

In this study we use only observations near the Galactic Anticentre, of fields with their centres in the region $160<l<200$, $-1<b<1$. Along sightlines towards the Galactic Anticentre Galactocentric radius increases at its greatest rate with heliocentric distance. As such, stars in the outer Galaxy are brightest along these sightlines. The Galactic Anticentre is a known node in the Galactic warp \citep[e.g.][]{Reyle_Marshall.2009}, so by choosing sightlines near this direction the influence of the warp on observations is reduced. 

It is possible to use the spatial distribution of a particular type or types of object to trace Galactic structure. Studies of the gaseous component of the disc have utilised the 21cm HI emission line \citep[e.g][]{Kalberla_Burton.2005}. To study the stellar density profile \cite{Robin_Creze.1992}, \cite{Ruphy_Robin.1996} and \cite{Reyle_Marshall.2009} have used photometry of all stars. This approach maximises the number of stellar tracers available, reducing statistical uncertainties. This was necessary given the relatively small samples available to \cite{Robin_Creze.1992} and \cite{Ruphy_Robin.1996}.

The cost of using all stars is that it conceals any distribution differences which occur between different types of star. Such contrasts could arise from changes in the structure and composition of the outer Galactic thin disc as a function of time. Even if a relatively restricted subset of stars is studied, a wide range of ages may be admitted and so results would be blurred. This is clearly true of subsets of late type-stars, which may cover an age range of up to 10~Gyrs.

In this study the tracers used are early-A type dwarfs. By selecting these the age range is minimised, as they only remain on the main sequence for $\sim 200$~Myrs. Therefore, we are measuring the distribution of stars that are relatively homogeneous with respect to age and so can avoid the problems discussed above for samples which span a broad range of ages. Early A-stars stars are also bright ($M_{r'}\sim1.5$) and so the distance range over which they can be studied is large. 

The IPHAS ($r'-~\Halpha$) colour is strongly correlated with the equivalent width of the $\Halpha$ line \citep{Drew_others.2005}. $\Halpha$ absorption is strongest in stars with effective temperatures of $\sim 9000~K$, which for solar metallicities corresponds to a spectral type of A2-A3. The stronger gravitational broadening of the main sequence stars, with respect to evolved stars, further strengthens the $\Halpha$ equivalent width and so separates these classes of stars on an ($r'-i',r'-\Halpha$) colour-colour diagram. This separation is not achieved with a purely broadband filter set. The effect of extinction on the IPHAS colour-colour diagram is such that there is good discrimination between reddening and intrinsic (i.e. unreddened) colour \citep{Drew_others.2005, Sale_others.2009}. Therefore, as described in detail by \cite{Drew_Greimel.2008}, early A-stars, on or near the main sequence, are easily identifiable in IPHAS colour-colour plots and it is straightforward to measure their extinction.

We adopt the colour cuts of \cite{Drew_Greimel.2008}, with $\delta ( r'~-~\Halpha)=0.03$. Solar metallicity stars selected in this manner will have spectral types in the approximate range A0--A5. Selected stars with super or sub-solar metallicities will formally match slightly different spectral types, but their range of intrinsic ($r'~-~i'$) colour and effective temperature will be unaffected. As only main sequence stars are included in the sample, the age range is relatively narrow, from roughly 10~Myr up to $\sim200$~Myr.

As observations approach the faint magnitude limit of the survey, the sample chosen will become progressively more mixed with other types of stars as a result of increasing photometric errors. Our simulations indicate, that, as expected, most of these contaminating stars are later A-type dwarfs. A magnitude limit of $r'\leq19$ was imposed on the sample, to keep the contamination level integrated across the entire sample below $\sim 10 \%$. Given an asymptotic Galactic extinction of $A_{r'}=2.5$, typical for low latitude sightlines near the Anticentre, and an absolute magnitude of $M_{r'}=1.5$, typical for stars in the sample \citep{Houk_Swift.1997}, this magnitude limit corresponds to a heliocentric distance range of 10~kpc. We found that altering the magnitude limit to $r'<19.5$ did not significantly affect the results obtained, despite increasing the overall contamination to $\sim 20\%$.

Initially, photometry obtained in observing conditions not meeting the survey threshold was removed from the prospective sample. Observations were rejected if they exhibited seeing $>1.7 \arcsec$, ellipticity $>0.2$ or a 10$\sigma$ limiting magnitude brighter than $r'=20$, which is indicative of low sky transparency. The seeing and ellipticity values used are those measured during the pipeline reduction of IPHAS data \citep[for more details see][]{Gonzalez-Solares_others.2008}. This cleaned sample contained observations of 271 fields. 

\begin{figure}
\centering
\includegraphics[width=80mm, height=60mm]{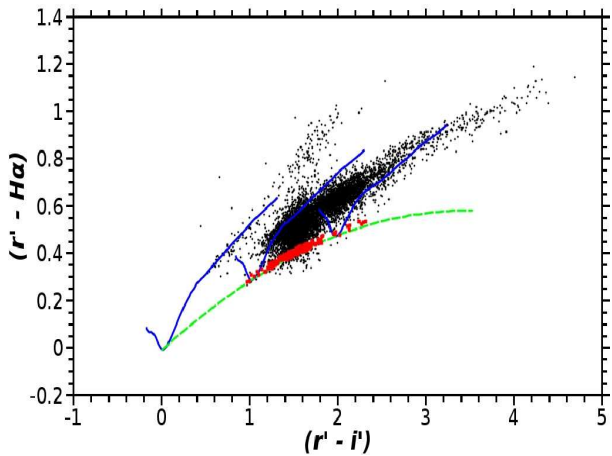}
\caption{Black points show data from IPHAS field 4199, where $13 \leq r' < 20$. Main sequences where extinctions equivalent to $A_{V}=0$, $4$ , $8$ for an A0V star have been applied are shown with solid blue lines and the dashed green line shows an A3V reddening line. Stars which would be selected by the colour-cut described above are shown with larger red points. \label{astars}}
\end{figure}

As the IPHAS survey has yet to be placed on a uniform photometric scale there is a minority of IPHAS observations which exhibit uncorrected photometric offsets. In such observations the loci of all stars, including  early A-stars, on the measured colour-colour plane shifts. \cite{Drew_others.2005} and Fig. \ref{astars} demonstrate that early A-stars always lie at the bottom of the main stellar locus. Therefore, their locus is easily observed and any offset from the proper position, as defined by synthetic spectral models, can be measured. Offsets in the $(r'-\Halpha )$ colour appear considerably more frequently than those in $(r'-i')$. Therefore, we characterise offsets by $\delta (r'-\Halpha )$, the distance in this colour between the locus of the early-A stars in the observed photometry and their proper position. In the unusual case of offsets only in $(r'-i')$ these will cause $\delta (r'-\Halpha )$ to be non-zero. Observations with a measured offset of $\mid \delta (r'-\Halpha ) \mid > 0.04$ were discarded, those with smaller offsets were corrected as necessary. The IPHAS observing includes a second set of exposures, displaced by $5\arcmin$ in both RA and Dec, taken immediately following the original set, it was possible to verify that objects observed in both sets of exposures had consistent photometry. For all fields where both sets of exposures were returned, the photometry was found to be consistent to well within the $2\%$ level in each band.

\begin{figure*}
\centering
\includegraphics[height=60mm]{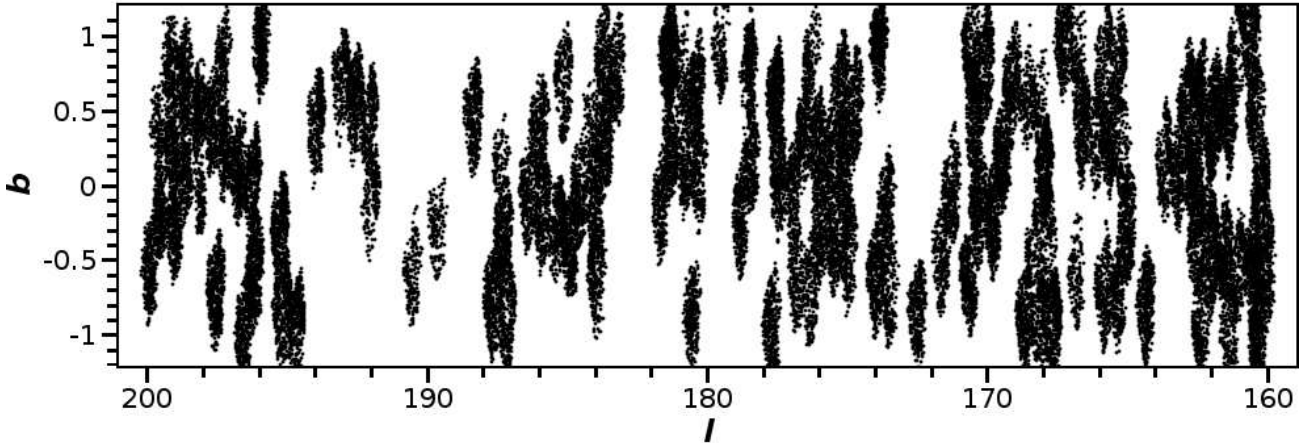}
\caption{The position of the 44,038 A-stars used in this study in galactic coordinates. \label{sky}}
\end{figure*}

The final sample contained observations of 132 fields and their offsets, in which there were 44,038 candidate early-A stars with $r'\leq19$. Fig. \ref{sky} shows the coverage of the final sample, which covers roughly 33 sq.deg. or $\sim 40\%$ of the total area from which we have drawn observations.

\section{Method}\label{method}

The general approach undertaken was to compare IPHAS observations to simulated photometry. The principle advantages of this method are that it allows us to take into account metallicity gradients and low level contamination of the sample of A-stars. Being able to deal with variations in metallicity is crucial as metallicity affects the absolute magnitude of stars. \cite{Hales_Barlow.2009} found that for a less conservative sample of A-stars, with $\delta \rm( r'-\Halpha \rm)$ up to 0.07, about $15\%$ of objects with $r'\leq18$ were not early A-stars. By properly accounting for this contamination the accuracy of the results obtained is improved. Our simulations indicate that the contamination is, at worst, $\sim 10 \%$. The simulation includes photometric errors and so simulated samples will contain stars with intrinsic colours outside the colour-cut used to select the sample, but which have been scattered in by photometric errors.

\cite{Robin_Creze.1992}, \cite{Ruphy_Robin.1996} and \cite{Reyle_Marshall.2009} have all undertaken a similar approach to studying the outer Galactic disc. By comparing photometry from the CFHT, DENIS and 2MASS respectively to a Galactic model they were able to constrain parameters including the thin disc scale length. However, in the first two cases the observations lack the quality and depth of more recent surveys. \cite{Reyle_Marshall.2009} cover a limited parameter space, apparently as a result of the large computational requirements involved in simulating a
Galactic model and determining extinction using this model and the method of \cite{Marshall_Robin.2006}.

\begin{figure}
\centering
\includegraphics[width=80mm, height=60mm]{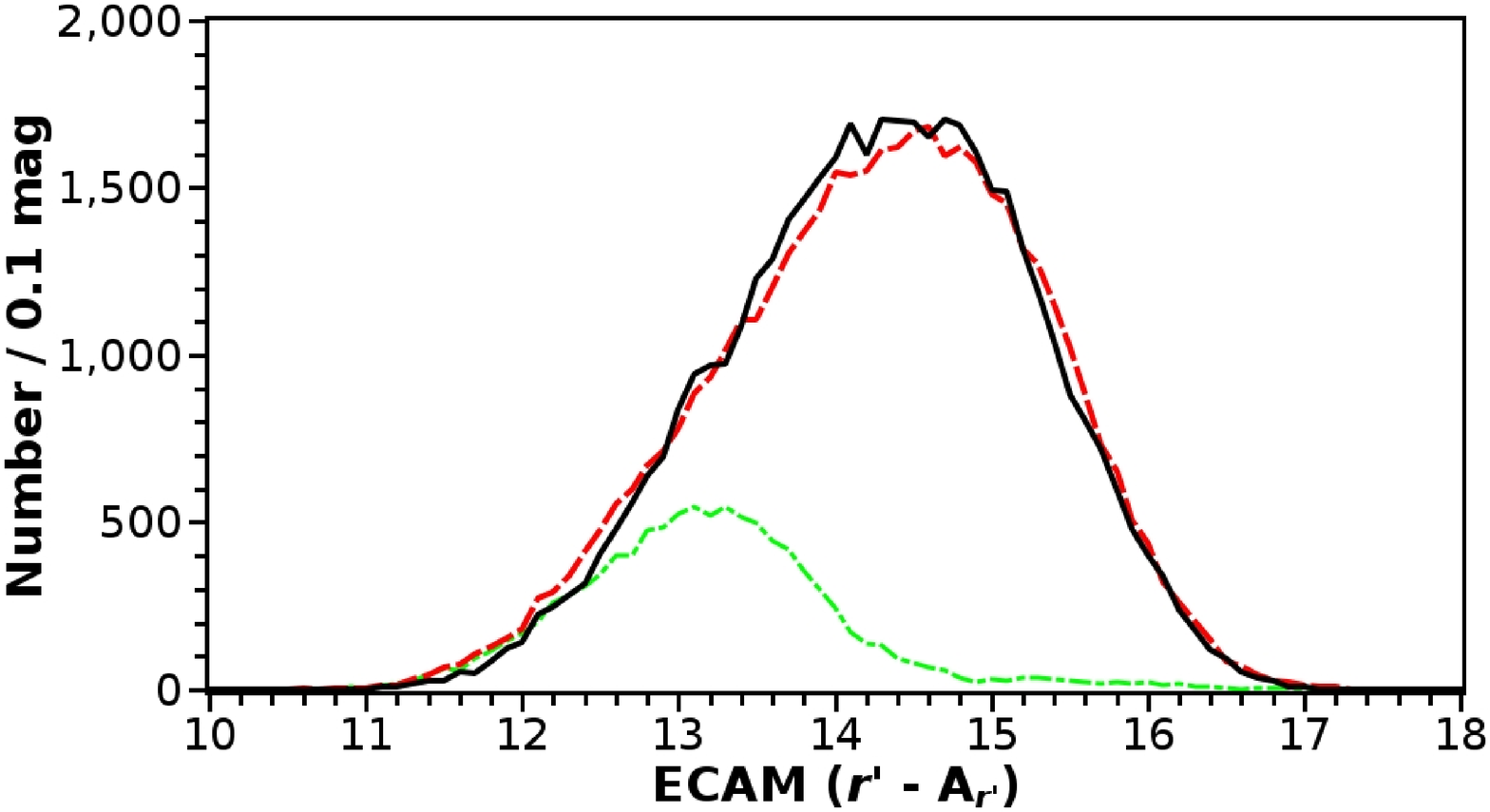}
\caption{Histograms of the ECAM distribution of the early A-stars. The actual observations are shown in black. The other two distributions are those of simulated photometry. The dashed red line indicates a model with an inner scale length of 2960~pc and an outer scale length of 1200~pc. The dot-dash green line one with an inner scale length of 2300~pc and an outer scale length of 0~pc.\label{chi-test}}
\end{figure}

As the reddening of a star can be determined from its position on the IPHAS colour-colour diagram, it is possible to correct the apparent magnitude of an object for this. This quantity can also be derived from the simulations. We refer to it as the extinction corrected apparent magnitude (ECAM) and is an estimate of the sum of the star's absolute magnitude and distance modulus. In this study ECAMs are always taken in the $r'$ band. The difference between the observed and simulated $r'$ ECAM distributions is quantified using a $\chi^2$ test. Our aim is to find the model parameter set which minimises this statistic. Figure~\ref{chi-test} demonstrates an example of poorly and better fitting models.

\subsection{Simulations of data}\label{Model}

The Galaxy model employed in this study uses many of the precepts of the Besan\c{c}on model of \cite{Robin_Reyle.2003}. It is similar to that employed in \cite{Sale_others.2009}. In this section we provide an outline of the model.

\begin{figure}
\centering
\includegraphics[width=80mm, height=60mm]{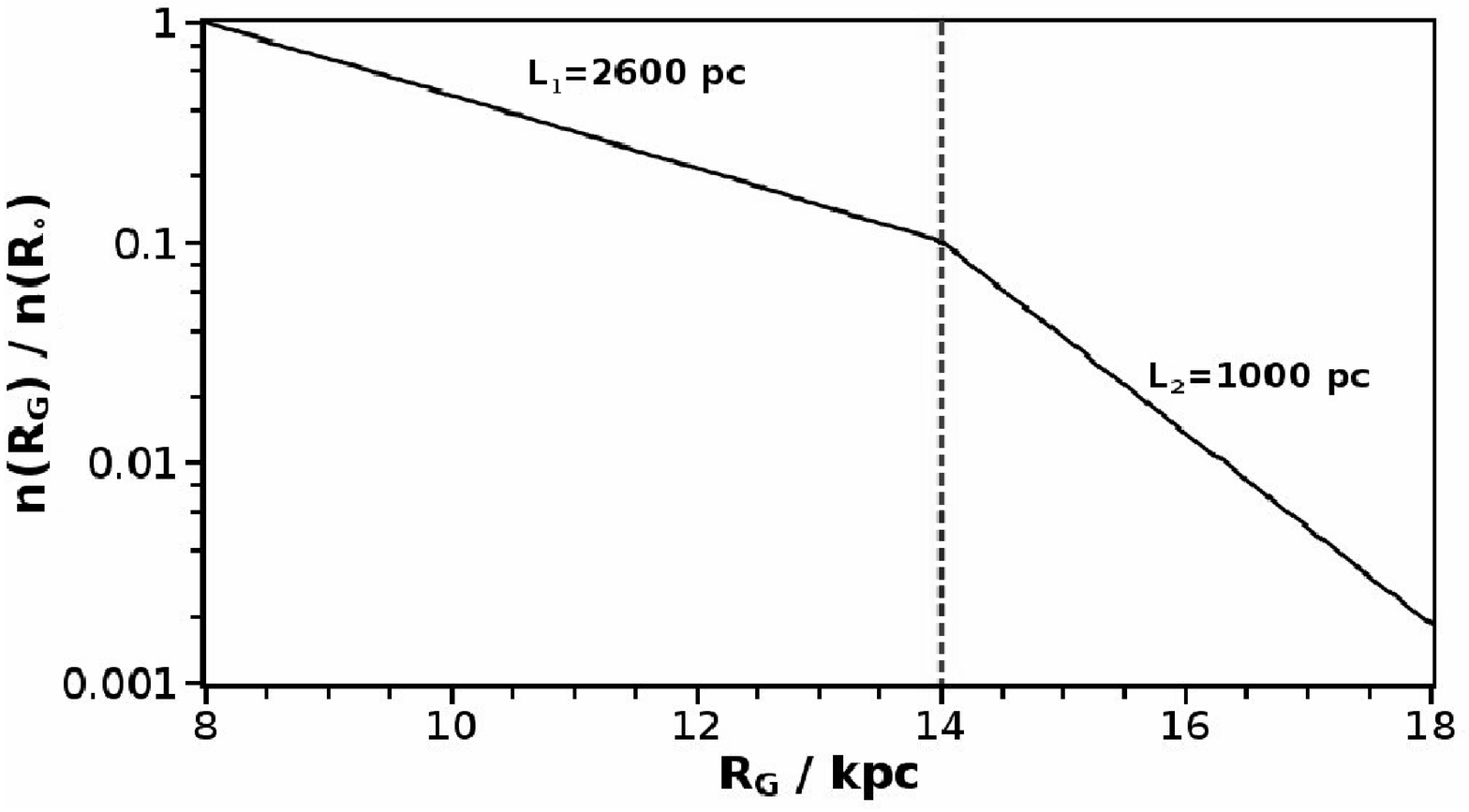}
\caption{A graphical depiction of the type of density profile employed in this study. The profile is defined by: the inner scale length, $L_1$; the truncation radius, $R_T$, which in this case occurs at $R_T=14$~kpc, as marked by the dashed vertical line; the outer scale length, $L_2$.  \label{density_profile}}
\end{figure} 

We define the disc's radial density profile with a double exponential; from the Solar Circle to some Galactocentric radius, here referred to as truncation radius ($R_T$), the density falls off with a scale length which we term the inner scale length ($L_1$). Beyond the truncation radius the density falls off with a different scale length, the outer scale length ($L_2$). Fig \ref{density_profile} demonstrates the form of the density profile we employ. We seek to measure these two scale lengths and the truncation radius and set constraints on the gradient of metallicity with respect to Galactocentric radius.

The stellar density laws for the thin disc, in the model we employ, are described as follows:

\begin{equation}
\rho(R,z) = \left\{ 
\begin{array}{l l}\rho_0/d_0 \times \exp(-(0.5^2+R^2/L_{1}^2+z^2/h^2)^{1/2}) & \quad \mbox{if $R \leq R_T$}\\
\rho_0/d_0 \times \exp(-(0.5^2+R^2/L_{2}^2+z^2/h^2)^{1/2}) & \quad \mbox{if $R > R_T$}\\ \end{array} \right. 
\end{equation}

\noindent Here $R$ is the Galactocentric radius, $z$ is the height above the Galactic plane, $\rho_0$ the local mass density, $d_0$ a normalisation factor and $\epsilon$ is the axis ratio. Values of $\rho_0$ are taken from \cite{Robin_Reyle.2003}.

The scale height, $h$ was set at 50~pc, similar to that implied by the Besan\c{c}on model. It is important to note that the scale height depends on the age of the populations studied; younger populations have lower scale heights. The adopted value of $h$ is in keeping with measurements of young objects as follows: \cite{Bonatto_Kerber.2006} find $48\pm3$~pc for open clusters with ages $<200$~Myrs, which is an equivalent age range to that of our study; \cite{Joshi_only.2007} find a scale height of $(57\pm4)$~pc for open clusters with ages $<300$~Myrs; \cite{Elias_Cabrera-Cano.2006} found $34\pm3$~pc for OB stars.

This study intentionally concentrates on the radial structure of the thin disc and is designed to be insensitive to the vertical structure. Using our simulations we were able to verify that reasonable variations of the scale height had no significant impact on the sample.

In this model the central hole in the disc, which is included in the original Besan\c{c}on model,  is neglected as it has no relevance beyond the Solar Circle. The thick disc was included, but as it is old ($\sim 11$~Gyrs) it has negligible influence on the results. Similarly, it is expected that including the stellar halo should have even less of a role and so it has not been included in simulations. 

We utilise the same description of warp and flare as in the Besan\c{c}on model. Warp is described by shifting the midplane by a distance $z_\mathrm{warp}$ at a Galactocentric radius of $R$ and a Galactic longitude of $l$, where:

\begin{equation}
z_\mathrm{warp} = z_c \sin (l -l_\mathrm{node})
\end{equation}

\begin{equation}
z_c = \gamma_\mathrm{warp} (R-R_\mathrm{warp})
\end{equation}

\noindent Here $l_\mathrm{warp}=180^{\circ}$, specifying a node in the Anticentre direction, $R_\mathrm{warp}=8.4$~kpc, with no warp within this radius, and $\gamma_\mathrm{warp}=0.18$. Similarly flare is described by increasing the scale height by a factor $k_\mathrm{flare}$, which is given by:

\begin{equation}
k_\mathrm{flare} = 1 + \gamma_\mathrm{flare}(R-R_\mathrm{flare})
\end{equation}

\noindent $\gamma_\mathrm{flare}$ takes the value $5.4  \times 10^{-2}$~kpc$^{-1}$ and $R_\mathrm{flare}$ is set at 9.5~kpc. As with warp, there is no flaring of the disc within this radius.

We assume the \cite{Scalo_only.1986} IMF, which is valid for field stars in the Galactic disc. The Star Formation Rate (SFR) used is taken from \cite{Robin_Reyle.2003}. The Galactic model assumes that neither the IMF nor the SFR vary with position in the thin disc. Therefore, these factors simply scale the number of A-stars in the simulations and we are not setting out to test them. This gives us the freedom to normalise the sample sizes of our simulated data sets as required and to confirm retrospectively that the implied IMF and SFR are reasonable.

The mean and variation of the metallicity of stars at the Solar Circle is taken from \cite{Robin_Reyle.2003}. In this description, stars in the solar neighbourhood with ages less than 1~Gyr have slightly super-solar metallicities. However, there is clear uncertainty on the mean metallicity of young stars in the solar neighbourhood, that must be borne in mind. A fuller discussion of this and its effects on the results we obtain is given in Section~\ref{metal_res}.

The most obvious alteration is that the model is adapted to return magnitudes in the IPHAS $r'i'\Halpha$ system, which are unsupported by the Besan\c{c}on model. The evolutionary tracks of \cite{Pietrinferni_Cassisi.2004} and \cite*{Siess_Dufour.2000} (for pre-main sequence objects) are used to convert the mass, age and metallicity of an object into $\rm T_{eff}, \log g$ and absolute magnitude. These are then converted into IPHAS colours and magnitudes using the \cite{Munari_Sordo.2005} grid of synthetic spectra, the IPHAS filter profiles and atmospheric and CCD response functions for the INT/WFC. 

Despite the fact that ECAMs are used in this study, it is still necessary to include extinction in the model as the photometric error of an object is, in part, dependent on the extinction it has suffered as well as on the distance to it. Rather than using a model of Galactic extinction, such as the models of \cite{Drimmel_Spergel.2001} or that used in the Besan\c{c}on model, extinction is determined empirically using the algorithm MEAD \citep{Sale_others.2009}. Given the extinction distance relationship found by MEAD, the monochromatic extinction of each star in the model is found based on its simulated heliocentric distances. The monochromatic extinction is converted into extinctions in each of the IPHAS bands using the $R=3.1$ extinction law of \cite{Fitzpatrick_only.2004}. It should be noted that this conversion is not quite linear with respect to the monochromatic extinction (particularly in the $\Halpha$ band) and is dependent on the SED of the source.

The growth of photometric errors as a function of apparent magnitude is taken from the observed data. The estimated photometric errors determined during pipeline processing of IPHAS data are fitted with a function of the following form:

\begin{equation}
\delta m = A + \exp^{(Bm+C)}
\label{error_growth}
\end{equation}

The effect of unresolved multiplicity in star systems was included in the model by adding a secondary star to 56\% of systems, following the binary fraction of \cite{Duquennoy_Mayor.1991}. The addition of secondary stars was done in such a way that the final single object IMF was consistent with the \cite{Scalo_only.1986} Galactic field IMF.

\subsection{Parameterisation of the simplified galactic model}\label{params-sec}

In this study we concentrated on four parameters; the gradient of metallicity with respect to Galactocentric radius and three governing the stellar density profile: inner scale length ($L_1$); outer scale length($L_2$); and the truncation radius ($R_T$), which is radius at which the scale length changes. 

Photometry was then simulated for 46,816 combinations of parameters, covering the following parameter ranges:

\begin{itemize}
\item Inner scale length ($L_1$): 19 values from 2300 to 3380 pc.
\item Outer scale length ($L_2$): 16 values from 0 to 3000pc
\item truncation radius ($R_T$): 14 values from 11 to 16.2 kpc.
\item Metallicity gradient($d[\mathrm{Fe}/ \mathrm{H}]/dR_G$): 11 values from -0.05 to -0.09~kpc$^{-1}$.
\end{itemize}

The model which best describes the observed data was found through the use of $\chi^2$ minimisation, comparing the ECAM distribution of the observed and model photometry. For each set of model photometry the sample size of the observed photometry was normalised to that of the model photometry. By taking this step it is only the relative distribution of ECAMs which is important and not the total number of objects. 

The number of early A-stars at the Solar Circle is a function of both the local stellar density and the IMF: steeper IMFs will result in fewer early A-stars. By rescaling the sample sizes our results are not dependant on the assumed local mass densities ($\rho_0$) or the IMF. 

Subsequently confidence limits were calculated on the four parameters, by simulating 20,000 more sets of photometry for the best fitting parameter set and then subsequently determining which of the original 46,816 models fitted it best, by the same process of $\chi^2$ minimisation. Note that the statistical uncertainties quoted in the following sections for $L_1, L_2, R_T$ and $d[\mathrm{Fe}/\mathrm{H}]/dR$ are the 68.3\% confidence intervals derived by this method.

\section{Results}\label{results}

\subsection{Summary of the best fitting parameters}

\begin{figure}
\centering
\includegraphics[width=80mm, height=60mm]{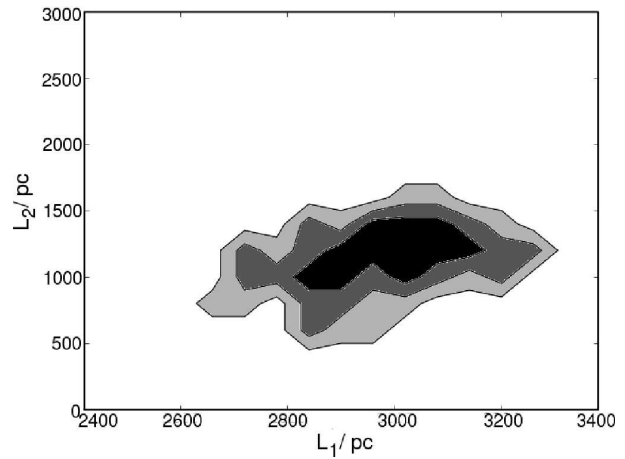}
\caption{A contour plot showing the derived confidence limits for the inner and outer scale lengths. Contours are at 68.3\%, 95.4\% and 99\%. Note that this plot is scaled to show the entire parameter range studied and that the range of $L_2$ shown is three times larger than that of $L_1$. \label{l1_l2}}
\end{figure}

An estimate of the inner scale length ($L_1$) of $3020$~pc with a 68.3\% confidence range of $\pm 120$~pc. The estimate of the inner scale length shows little correlation with the estimates of the other parameters (e.g. Fig.~\ref{l1_l2}).

The metallicity gradient found is $(-0.07 \pm 0.01 ){\rm kpc}^{-1}$. This is less tightly constrained, but is well within the range of estimates in the literature \citep[see e.g.][]{Lemasle_Piersimoni.2008}.

\begin{figure}
\centering
\includegraphics[width=80mm, height=60mm]{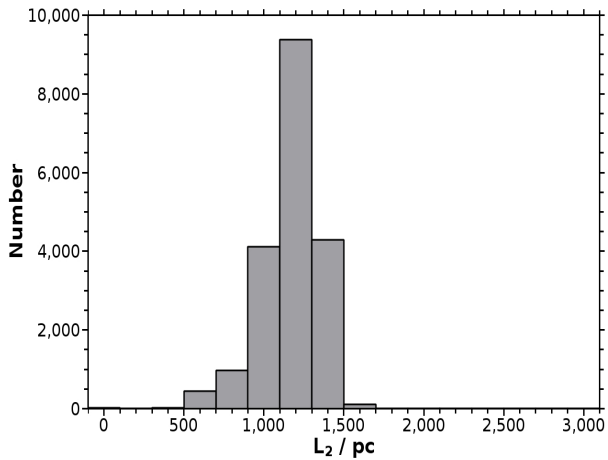}
\caption{Plot showing the distribution of the 20,000 visualisations used to determine the confidence intervals with respect to the outer scale length. \label{s2_only}}
\end{figure}

\begin{figure}
\centering
\includegraphics[width=80mm, height=60mm]{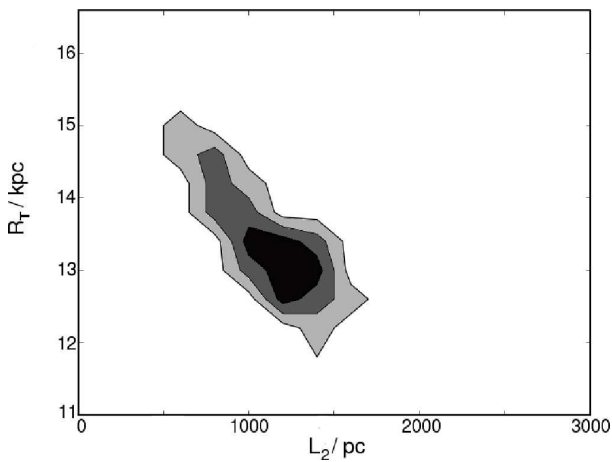}
\caption{A contour plot showing the derived confidence limits on outer scale length and truncation radius. Contours are at 68.3\%, 95.4\% and 99\%.  Note that this plot is scaled to show the entire parameter range studied. \label{s2_cr}}
\end{figure}

\begin{figure}
\centering
\includegraphics[width=80mm, height=60mm]{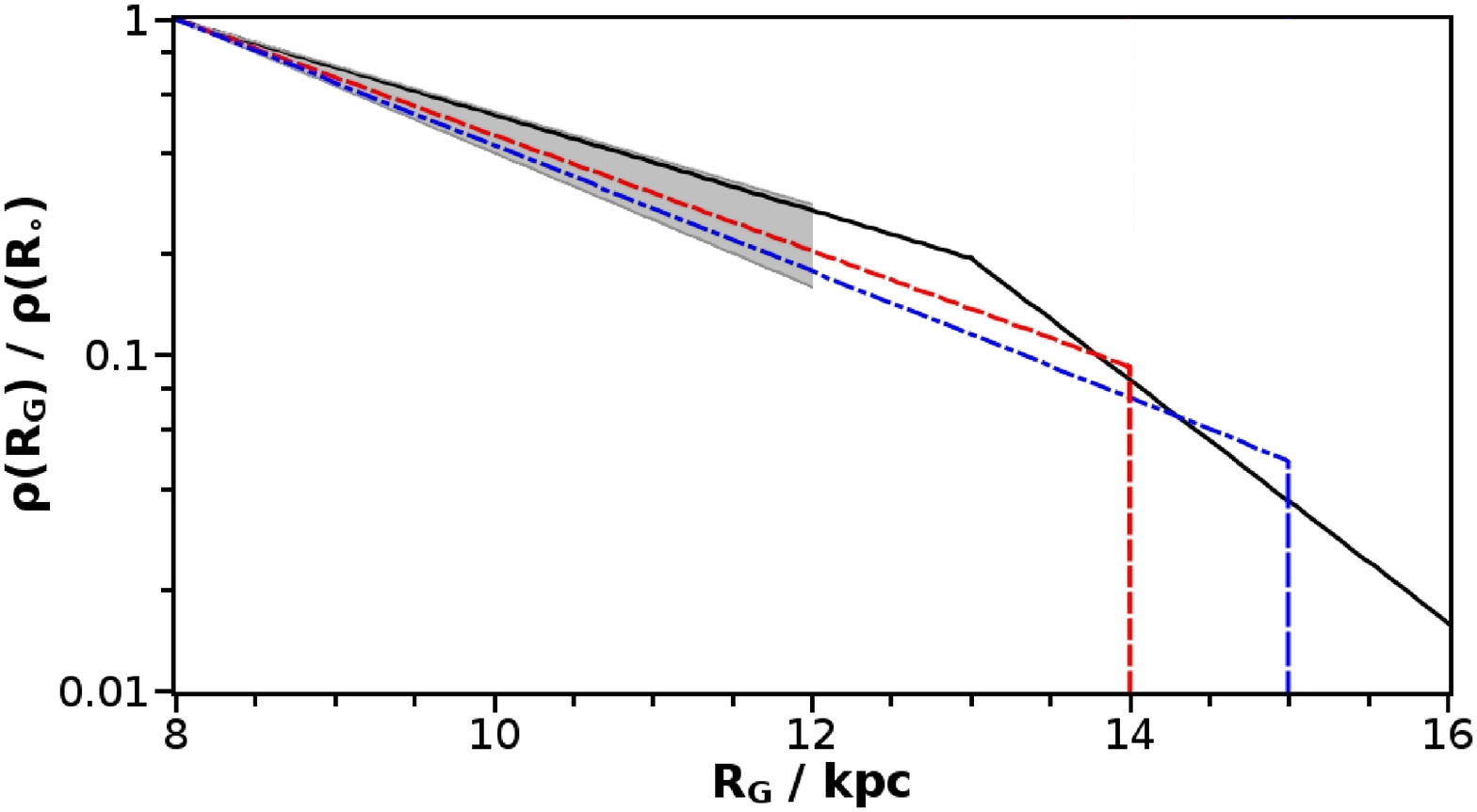}
\caption{The best fitting radial density profile derived in this study is shown in black. For comparison, the profiles of \protect\cite{Robin_Creze.1992} (red dashed line) and \protect\cite{Ruphy_Robin.1996} (blue dot dashed line) are also shown, whilst the grey region indicates the inner scale length found by \protect\cite{Juric_others.2008}. Note that the high Galactic latitude of the SDSS observations employed by \protect\cite{Juric_others.2008} prevents them observing features far beyond the Solar Circle and so they cannot observe the truncation of the disc. All profiles on this plot have been normalised to the density at the Solar Circle. \label{radial_prof}}
\end{figure}

We have been able to determine the outer scale length, $L_2$, to be $(1200 \pm 300)$~pc. The probability of the length scale being zero is very small, $\sim 0.05 \%$, see Fig. \ref{s2_only}. The truncation radius is estimated at $R_T=(13.0 \pm 0.5)$~kpc. The estimates of these two qualities are strongly covariant, see Fig. \ref{s2_cr}, so that a longer estimate of $L_2$ would imply a shorter $R_T$. The radial density profile derived is compared to others from the literature in figure \ref{radial_prof}. 

\subsection{Inner scale length}

As discussed in Section \ref{Model} our approach to dealing with stellar multiplicity is to account for it from the outset. This contrasts with the approaches adopted in previous studies \citep{Siegel_Majewski.2002, Juric_others.2008}, where the scale length measurement assumes that all sources are single stars. These earlier authors preferred to apply an a posteriori correction to account for binarity. In particular \cite{Siegel_Majewski.2002} simulated photometry, with varying binary fraction, and then remeasured the scale lengths in the same manner as they had done for the real data. As the scale lengths in the simulations were known, the ratio between these and the lengths measured provided the required corrections, as a function of binary fraction. Given a binary fraction of $50\%$, \cite{Siegel_Majewski.2002} estimate that their measured scale lengths will be only $\sim80\%$ of the true value. They then go on to correct their results accordingly. \cite{Juric_others.2008} adopt the same correction.

As we have included binaries in our models, our results do not require a correction of this form. However, to investigate the impact of assuming binarity, we have obtained the best fitting inner scale length using models without binarity. In order to focus on the inner scale length, we imposed a magnitude limit of $r'<16.5$, so that only stars expected to lie within $R_T$ are included. We found that the estimated values of the inner scale length using models without binarity were $(79 \pm 2)\%$ of the values found using models with binarity. This factor is in close agreement with that determined by \cite{Siegel_Majewski.2002} and subsequently reused by \cite{Juric_others.2008}. 

If we apply this factor to our results, effectively carrying out the reverse of the correction of \cite{Siegel_Majewski.2002}, we obtain a scale length of $(2400 \pm 80)$~pc, which is longer than the uncorrected scale lengths of both \cite{Siegel_Majewski.2002} and \cite{Juric_others.2008}.

\subsection{Outer Scale length and truncation radius}

Our estimate of $R_T=(13.0 \pm 0.5$)~kpc is substantially shorter than that of \cite{Robin_Creze.1992} (14~kpc) and \cite{Ruphy_Robin.1996} ($15 \pm 2$~kpc). Their parameterisation of the radial stellar density profile was simpler than ours, in adopting complete truncation of the thin disc ($L_2=0$~pc), as seen in Fig.~\ref{radial_prof}. We find that our estimates of $R_T$ and $L_2$ are anti-correlated (Fig \ref{s2_cr}). So that if we were to assume $L_2=0$~pc, the best fitting distribution would be with $R_T=15.5$~kpc, in agreement with the estimate of \cite{Ruphy_Robin.1996}. This suggests that the simpler formulation assumed by \cite{Robin_Creze.1992} and \cite{Ruphy_Robin.1996} led to the substantially longer estimate of $R_T$.

\subsection{Metallicity}\label{metal_res}

It has been proposed that the Galactic metallicity profile could be described as a step function of Galactocentric radius, with a break at a radius of $\sim 10$~kpc \citep{Corder_Twarog.2001, Yong_Carney.2005}. For our sample, such a metallicity distribution would result in there being a range of apparent magnitudes over which there are few early A-dwarfs. This is because those dwarfs immediately beyond the break would have discontinuously fainter absolute magnitudes than those in front, as a result of their lower metallicity. Such a distribution was tested for a range of stellar density profiles and was on average $\sim 100$ times less likely than a constant metallicity gradient of $-0.070$~kpc$^{-1}$.

A further suggested metallicity profile is one where the metallicity profile flattens in the outer Galaxy. \cite{Carraro_Geisler.2007}, \cite{Bragaglia_Sestito.2008} and \cite{Lemasle_Piersimoni.2008} observe such a profile, whilst \cite{Chiappini_Matteucci.1997} and \cite{Cescutti_Matteucci.2007} find it to be the natural result of their models of the chemodynamical evolution of a Milky Way like galaxy. In practice, the difference between such a metallicity profile and one with a fixed metallicity gradient is only manifested in small changes in absolute magnitude for stars at large Galactocentric radii, close to and beyond the reach of our observations. As a result, the method described here is not able to discriminate between these two profiles.

The mean metallicity of young stars in the solar neighbourhood is uncertain. Empirical age-metallicity relationships for the solar neighbourhood do not typically extend to populations as young as $\sim 100$~Myrs \citep{Haywood_only.2006, Holmberg_Nordstr"om.2009}. \cite{Chen_Hou.2003} measured the metallicity and distances of a catalogue of 119 open clusters. From these we make a subsample of nine clusters, which lie within 500~pc of the sun amd 50~pc of the Galactic midplane. This sample has a mean metallicity of $[\mathrm{Fe}/\mathrm{H}]=-0.01$. \cite{Viana_Almeida_Santos.2009} found that the stars in 11 nearby young associations have an average metallicity of $[\mathrm{Fe}/\mathrm{H}]=0.00$. \cite{Viana_Almeida_Santos.2009} also perform a correction to the similar data of \cite{Santos_Melo.2008} determining a mean value of $[\mathrm{Fe}/\mathrm{H}]=0.06$ for a smaller sample of stars drawn from six local associations.

We conservatively adopt an estimated uncertainty of 0.1~dex on the mean metallicity of young stars in the solar neighbourhood. The results listed above would formally indicate a smaller uncertainty, but our broader measure allows for systematic error in the determination of those values. Based on the isochrones of \cite{Pietrinferni_Cassisi.2004}, a shift of 0.1~dex in metallicity would imply an alteration of 0.06 to the $r'$ band absolute magnitude of early A-dwarfs. Such an alteration would shrink or expand the distance to all the stars in the sample by 6\% and so alter the measurements of $L_1, L_2$ and $R_T-R_{\odot}$ in the same proportion. Therefore, we include this as an additional source of uncertainty on our estimates of these quantities, with values given in Table~\ref{params}.

\subsection{Galactocentric radius of the Sun}

There is some disagreement on the distance of the Sun from the Galactic centre ($R_{\odot}$). In the study we have adopted the solar Galactocentric radius of 8.07~kpc from \cite{Trippe_Gillessen.2008}. However, there is substantial uncertainty on this estimate: 0.32~kpc of statistical uncertainty and 0.17~kpc of systematic error. In addition, other recent estimates show some scatter: \cite{Ghez_Salim.2008} found ($8.0\pm0.6$)~kpc, whilst \cite{Gillessen_Eisenhauer.2009} found ($8.33\pm0.35$)~kpc, by reanalysing the data of \cite{Eisenhauer_others.2005}. Given this dispersion, simulated photometry was created for a small subset of the original set of parameters, with $R_{\odot}$ taking the values of 7.57, 7.82, 8.32 and 8.57~kpc. As with the test for the effect of binarity, a magnitude limit of $r'<16.5$ was imposed.  The values of $L_1$ obtained varied by up to $2\%$ around the estimate with the solar Galactocentric radius of 8.07~kpc. Therefore, we include this as an additional source of error. The effect of the uncertainty on the value of $R_{\odot}$ on our estimates of the values of the inner scale length and truncation radius is listed in table~\ref{params}. The metallicity gradient and the outer scale length are relatively poorly constrained, Correspondingly, the influence of the modest uncertainty in the solar Galactocentric radius on these values is insignificant.

\subsection{Consistency of results with respect to Galactic longitude}

Although the values of the parameters obtained are the best fitting for the entire observed data set, it is not clear that this is true when trends with Galactic longitude are considered. For these reasons we split all our data (both simulations and observations) into 3 longitude subsets ($160\leq l <170; 175 \leq l <185; 190 \leq l <200$) and again applied the brighter magnitude limit of $r'<16.5$, to focus on how our estimate of $L_1$ varies with sightline. Then, the inner scale length was determined with each sample. We found inner scale lengths of: $(3030 \pm 320)$~pc, $(2960 \pm 290)$~pc and $(3140 \pm 330)$~pc respectively for the three samples. These results are consistent with each other and the result for the entire sample. 

\subsection{Other parameters and features influencing results}

In addition to the four parameters discussed above, there are several other parameters which can be varied in the model. These include the shape of the IMF, the form of any warp and flare of the Galactic disc and the shape of the radial abundance profile. The effect of varying these parameters was addressed on some of the 46,816 parameter sets tested above. 

\cite{Reyle_Marshall.2009} measure the magnitude of the Galactic warp to be somewhat smaller than that assumed here. However, by design, the form of the warp had negligible impact on the data, we selected the direct Anticentre for study precisely because evidence suggests it contains a node in the Galactic warp \citep{Reyle_Marshall.2009}.

The effect of flare in the outer Galactic thin disc is to redistribute stars to greater heights above and below the midplane. This in turn will affect observations in two ways: the distance to the stars will be altered and the number which lie within a given lattitude range will drop. As the intrinsic latitude range of our A-stars is small, the change in distance will be very small and so will have no significant affect on the observed ECAMs. Therefore, flare can only affect our results by reducing the number of objects captured within the latitude range we study. Using our simulations we have calculated that removing the flare increases the sample size by only $\sim 200$ objects ($\sim 0.5 \%$), almost all inside $R_T$. This is a result of the relatively low scale height of the young populations from which our A-stars are drawn (see Section~\ref{Model}). So, it is expected that only a dramatic increase in the flaring of the outer disc could significantly affect our results. For the same reason, the radial truncation of the thin disc that we observe cannot be the result of an incorrect treatment of the disc's flare.

As discussed in section~\ref{Model}, by assumption the SFR and the shape of the single star IMF affects the number of stars in the early A range,  but without altering their distribution with respect to Galactocentric radius. We have conducted simulations to confirm that the key parameters are unaffected by varying the IMF slope from $\alpha=-2$ to $-3$, as expected.

\section{Conclusions}\label{conclusions}

\begin{table*}
\caption{The best fitting values of the parameters, with their associated confidence intervals. \label{params}}
\begin{tabular}{ c c c c c c }
 Parameter & Best fitting & 68\% Confidence & Uncertainty Resulting From 0.1 dex & Uncertainty Resulting & Total Systematic\\
  & Value & Limit &  Uncertainty On Local Mean Metallicity & From Estimate Of $R_{\odot}$  & Uncertainty \\ 
\hline
$L_1$ & 3020~pc & $\pm 120$~pc & $\pm 170$~pc & $\pm 50$~pc & $\pm180$~pc\\
$L_2$ & 1200~pc & $\pm 300$~pc & $\pm 70$~pc & -- & $\pm70$~pc \\
$R_T$ & 13~kpc & $\pm 0.5$~kpc & $\pm 0.3$~kpc & $\pm 0.5$~kpc & $\pm0.6$~kpc\\
$\frac{d[\mathrm{Fe}/\mathrm{H}]}{R_G}$ & -0.07~kpc$^{-1}$ & $\pm 0.01$~kpc$^{-1}$ & -- & -- & --\\
\end{tabular}
\end{table*}

The values of the parameters derived in this study are summarised in Table~\ref{params}. Also listed are the uncertainties on these values, where the systematic component is the combination of those due to the uncertainty in $R_{\odot}$ and in the local mean metallicity. For the reasons given in Section~\ref{metal_res}, we have adopted a conservative estimate of the latter. As a result the uncertainties the total systematic error, to which it contributes, is similarly generous.

The inner scale length of the thin disc measured in this study is $(3020 \pm 120_{statistical} \pm 180_{systematic})$~pc. \cite{Siegel_Majewski.2002} found $(2500-3125)$~pc and \cite{Juric_others.2008} $(2600 \pm 650)$~pc. Fig.~\ref{radial_prof} summarises some of the different estimates obtained. Both \cite{Siegel_Majewski.2002} and \cite{Juric_others.2008} utilise samples of stars which have large mean ages. The measurement of \cite{Juric_others.2008} is largely based on K and M dwarfs, which will have a mean age of several Gyrs. This contrasts with our sample of stars, which is predominantly young, early A-stars will only spend $\sim 200$~Myrs on the main sequence. If the Galaxy was undergoing inside--out formation, it would be expected that the inner scale length measured for our sample would be longer than those obtained from the predominantly older samples employed in other studies. The results obtained are consistent with this scenario, though statistical and systematic uncertainties prevent any direct conclusions being drawn at this stage.

Additionally, we have utilised observations taken very near the Galactic plane, thus maximising our sensitivity to the density profile of the thin disc. Our estimate of $L_1$ is longer than those of \cite{Robin_Creze.1992} and \cite{Ruphy_Robin.1996}, but it is unclear how these two studies have treated binarity. 

In common with \cite{Robin_Creze.1992} and \cite{Ruphy_Robin.1996} we find that there is a knee in the radial density profile. However, where as they proposed that the disc is completely truncated at that point, we find evidence for the disc continuing, albeit with shorter scale length. The more gradual truncation that we observe would appear to be more physically reasonable, given that complete truncation would rule out both low level star formation outside the truncation radius \citep[discussed in][]{Elmegreen_Hunter.2006} and the outwards radial migration of stars beyond the truncation radius, as seen in the model of \cite{Rovskar_Debattista.2008}.

Given our findings, the Galaxy appears to be an example of a Type II galaxy on the scheme proposed by \cite{Freeman_only.1970}. Such galaxies appear to be common in the Universe, accounting for $60\%$ of the sample of \cite{Pohlen_Trujillo.2006}. We find that the ratio of the inner to outer scale lengths to be $2.5 \pm 0.6$, consistent with the mean ratio of $2.1\pm0.5$ found by \cite{Pohlen_Trujillo.2006} for classically truncated disc galaxies.

\cite{Naab_Ostriker.2006} produced a Galactic model, incorporating many observed properties of the Milky Way. They tested the effect of star formation thresholds on their model and concluded that they would truncate the stellar surface density at a Galactocentric radius of $\sim 12$~kpc. This result appears to be shorter than the truncation radius we have measured.

We have measured a metallicity gradient of $(-0.07 \pm 0.01 ){\rm kpc}^{-1}$, which is within the range of previous measurements in the literature \citep[see e.g.][]{Lemasle_Piersimoni.2008}. Frequently, the ages of the tracers used in previous studies have covered a wide range of ages, compared to our early A-stars which are between 10 and 200~Myrs old. Therefore, our results are indicative of the metallicity of the ISM in the recent past, whereas other studies, such as \cite{Bragaglia_Tosi.2006}, have utilised older tracers, which thus probe the ISM at a different point in the history of the Galaxy. Our approach contrasts with previous studies, in that instead of using a small number of objects, each of which have well measured metallicities, we have utilised the characteristics of a large population to determine a metallicity profile. We have also managed to rule out the possibility of the radial metallicity profile following a step function \cite[as sugested by][ and others]{Yong_Carney.2005}.

\begin{figure*}
\centering
\includegraphics[width=160mm]{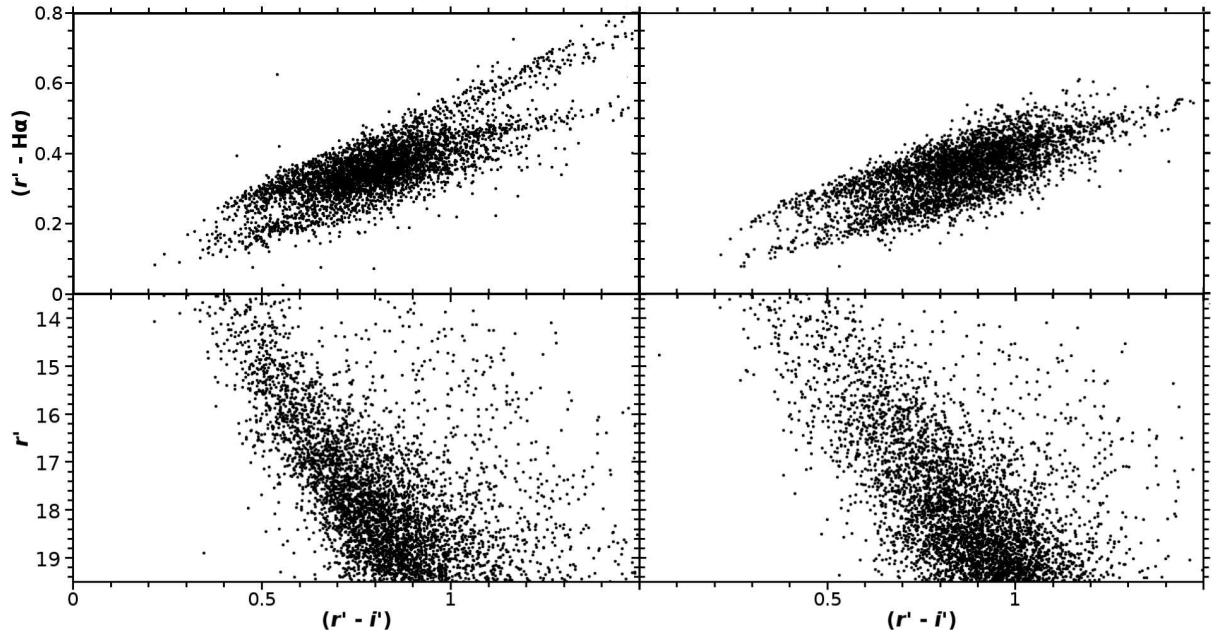}
\caption{Actual (left) and simulated (right) IPHAS colour-colour (top) and colour-magnitude (bottom) diagrams for IPHAS field 2566, with its field centre in the direction (l,b)=(175.6,0.2). The total area of the field is 0.28 sq.deg. The simulations lack M-dwarfs, which exhibit large $(r'-i')$ and $(r'-\Halpha)$ colours and more exotic objects such as $\Halpha$ emitters, of which there are three prominent examples in this field.\label{compare}}
\end{figure*} 

Fig.~\ref{compare} compares observed and simulated IPHAS data for the same location on the sky, where the growth of photometric errors in the simulations have been copied from those in the observations and the extinction in the simulations has been derived from the observations with MEAD. The differences between the real and simulated observations arise for several reasons. The simulated photometry lacks low mass ($M<0.5M_{\odot}$) stars and so does not exhibit the large number of M-dwarfs visible in the observations, with large $(r'-i')$ and $(r'-\Halpha)$ colours. Although the relative distribution of early A-dwarfs is not dependant on factors such as the IMF and SFR, the appearance of both the colour-colour and colour-magnitude planes is. In the future studying the distribution of stars drawn from a broader spectral type range should allow these factors to be estimated. Our simulated photometry has been derived from a model of a smooth axisymmetric Galaxy. On the larger scales considered by this study such an approximation is sound, but on the smaller scales, such as that of an indicative IPHAS field (Fig.~\ref{compare}), it is not. Future use of the IPHAS data will include a full mapping of the outer Galactic disc with all available IPHAS data over the range $b<5^{\circ}$. This  will enable departures from smoothness to be clearly observed and lead to more focus on the vertical structure.

This study has only utilised a small fraction of the available IPHAS data, with the sample limited by both spectral type and area on the sky. Extension to later spectral types may allow the variation of the density profile with stellar ages to be studied, possibly allowing the inside--out formation of the Galaxy to be studied. It is currently intended to produce both observed stellar density and extinction maps for the entirety of this region. Additionally, there are several other ongoing (UKIDSS-GPS \citealt{Lucas_others.2008}, UVEX \citealt{Groot_others.2009}) and upcoming (VPHAS+, VVV) optical and near infra-red surveys of the Galactic plane, the observations of which could also be used to further analyse the Galactic thin disc.

\section*{Acknowledgements}

The authors would like to thank the referee, Annie Robin, for her comments that have helped improve this paper. The authors would also like to thank Kerttu Viironen for her helpful comments.

This paper makes use of data obtained as part of the INT Photometric H$\alpha$ Survey of the northern Galactic Plane (IPHAS) carried out at the Isaac Newton Telescope (INT). The INT is operated on the island of La Palma by the Isaac Newton Group in the Spanish Observatorio del Roque de los Muchachos of the Instituto de Astrofisica de Canarias. All IPHAS data are processed by the Cambridge Astronomical Survey Unit, at the Institute of Astronomy in Cambridge. SES is in receipt of a studentship funded by the Science \& Technology Facilities Council of the United Kingdom.

The simulations presented here have been prepared using the Imperial College High Performance Computing Service.

\bibliography{main-2}

\begin{thebibliography}{}

\bibitem[\protect\citeauthoryear{{Azzollini}, {Trujillo} \&
  {Beckman}}{{Azzollini} et~al.}{2008}]{Azzollini_Trujillo.2008}
{Azzollini} R.,  {Trujillo} I.,    {Beckman} J.~E.,  2008, \apj, 684, 1026

\bibitem[\protect\citeauthoryear{{Bakos}, {Trujillo} \& {Pohlen}}{{Bakos}
  et~al.}{2008}]{Bakos_Trujillo.2008}
{Bakos} J.,  {Trujillo} I.,    {Pohlen} M.,  2008, \apjl, 683, L103

\bibitem[\protect\citeauthoryear{{Barker}, {Sarajedini}, {Geisler}, {Harding}
  \& {Schommer}}{{Barker} et~al.}{2007}]{Barker_Sarajedini.2007}
{Barker} M.~K.,  {Sarajedini} A.,  {Geisler} D.,  {Harding} P.,    {Schommer}
  R.,  2007, \aj, 133, 1125

\bibitem[\protect\citeauthoryear{{Battaner}, {Florido} \&
  {Jim{\'e}nez-Vicente}}{{Battaner} et~al.}{2002}]{Battaner_Florido.2002}
{Battaner} E.,  {Florido} E.,    {Jim{\'e}nez-Vicente} J.,  2002, \aap, 388,
  213

\bibitem[\protect\citeauthoryear{{Bland-Hawthorn}, {Vlaji{\'c}}, {Freeman} \&
  {Draine}}{{Bland-Hawthorn} et~al.}{2005}]{Bland-Hawthorn_Vlajic.2005}
{Bland-Hawthorn} J.,  {Vlaji{\'c}} M.,  {Freeman} K.~C.,    {Draine} B.~T.,
  2005, \apj, 629, 239

\bibitem[\protect\citeauthoryear{{Boissier}, {Prantzos}, {Boselli} \&
  {Gavazzi}}{{Boissier} et~al.}{2003}]{Boissier_Prantzos.2003}
{Boissier} S.,  {Prantzos} N.,  {Boselli} A.,    {Gavazzi} G.,  2003, \mnras,
  346, 1215

\bibitem[\protect\citeauthoryear{{Bonatto}, {Kerber}, {Bica} \&
  {Santiago}}{{Bonatto} et~al.}{2006}]{Bonatto_Kerber.2006}
{Bonatto} C.,  {Kerber} L.~O.,  {Bica} E.,    {Santiago} B.~X.,  2006, \aap,
  446, 121

\bibitem[\protect\citeauthoryear{{Bragaglia}, {Sestito}, {Villanova},
  {Carretta}, {Randich} \& {Tosi}}{{Bragaglia}
  et~al.}{2008}]{Bragaglia_Sestito.2008}
{Bragaglia} A.,  {Sestito} P.,  {Villanova} S.,  {Carretta} E.,  {Randich} S.,
    {Tosi} M.,  2008, \aap, 480, 79

\bibitem[\protect\citeauthoryear{{Bragaglia} \& {Tosi}}{{Bragaglia} \&
  {Tosi}}{2006}]{Bragaglia_Tosi.2006}
{Bragaglia} A.,  {Tosi} M.,  2006, \aj, 131, 1544

\bibitem[\protect\citeauthoryear{{Bresolin}, {Ryan-Weber}, {Kennicutt} \&
  {Goddard}}{{Bresolin} et~al.}{2009}]{Bresolin_Ryan-Weber.2009}
{Bresolin} F.,  {Ryan-Weber} E.,  {Kennicutt} R.~C.,    {Goddard} Q.,  2009,
  \apj, 695, 580

\bibitem[\protect\citeauthoryear{{Buser}, {Rong} \& {Karaali}}{{Buser}
  et~al.}{1998}]{Buser_Rong.1998}
{Buser} R.,  {Rong} J.,    {Karaali} S.,  1998, \aap, 331, 934

\bibitem[\protect\citeauthoryear{{Carraro}, {Geisler}, {Villanova},
  {Frinchaboy} \& {Majewski}}{{Carraro} et~al.}{2007}]{Carraro_Geisler.2007}
{Carraro} G.,  {Geisler} D.,  {Villanova} S.,  {Frinchaboy} P.~M.,
  {Majewski} S.~R.,  2007, \aap, 476, 217

\bibitem[\protect\citeauthoryear{{Cescutti}, {Matteucci}, {Fran{\c c}ois} \&
  {Chiappini}}{{Cescutti} et~al.}{2007}]{Cescutti_Matteucci.2007}
{Cescutti} G.,  {Matteucci} F.,  {Fran{\c c}ois} P.,    {Chiappini} C.,  2007,
  \aap, 462, 943

\bibitem[\protect\citeauthoryear{{Chen}, {Hou} \& {Wang}}{{Chen}
  et~al.}{2003}]{Chen_Hou.2003}
{Chen} L.,  {Hou} J.~L.,    {Wang} J.~J.,  2003, \aj, 125, 1397

\bibitem[\protect\citeauthoryear{{Chiappini}, {Matteucci} \&
  {Gratton}}{{Chiappini} et~al.}{1997}]{Chiappini_Matteucci.1997}
{Chiappini} C.,  {Matteucci} F.,    {Gratton} R.,  1997, \apj, 477, 765

\bibitem[\protect\citeauthoryear{{Colavitti}, {Cescutti}, {Matteucci} \&
  {Murante}}{{Colavitti} et~al.}{2009}]{Colavitti_Cescutti.2009}
{Colavitti} E.,  {Cescutti} G.,  {Matteucci} F.,    {Murante} G.,  2009, \aap,
  496, 429

\bibitem[\protect\citeauthoryear{{Corder} \& {Twarog}}{{Corder} \&
  {Twarog}}{2001}]{Corder_Twarog.2001}
{Corder} S.,  {Twarog} B.~A.,  2001, \aj, 122, 895

\bibitem[\protect\citeauthoryear{{Daflon} \& {Cunha}}{{Daflon} \&
  {Cunha}}{2004}]{Daflon_Cunha.2004}
{Daflon} S.,  {Cunha} K.,  2004, \apj, 617, 1115

\bibitem[\protect\citeauthoryear{{de Vaucouleurs} \& {Pence}}{{de Vaucouleurs}
  \& {Pence}}{1978}]{deVaucouleurs_Pence.1978}
{de Vaucouleurs} G.,  {Pence} W.~D.,  1978, \aj, 83, 1163

\bibitem[\protect\citeauthoryear{Drew et~al.,}{Drew
  et~al.}{2005}]{Drew_others.2005}
Drew J.~E.,  et~al., 2005, \mnras, 362, 753

\bibitem[\protect\citeauthoryear{{Drew}, {Greimel}, {Irwin} \& {Sale}}{{Drew}
  et~al.}{2008}]{Drew_Greimel.2008}
{Drew} J.~E.,  {Greimel} R.,  {Irwin} M.~J.,    {Sale} S.~E.,  2008, \mnras,
  386, 1761

\bibitem[\protect\citeauthoryear{{Drimmel} \& {Spergel}}{{Drimmel} \&
  {Spergel}}{2001}]{Drimmel_Spergel.2001}
{Drimmel} R.,  {Spergel} D.~N.,  2001, \apj, 556, 181

\bibitem[\protect\citeauthoryear{{Duquennoy} \& {Mayor}}{{Duquennoy} \&
  {Mayor}}{1991}]{Duquennoy_Mayor.1991}
{Duquennoy} A.,  {Mayor} M.,  1991, \aap, 248, 485

\bibitem[\protect\citeauthoryear{{Eisenhauer} et~al.,}{{Eisenhauer}
  et~al.}{2005}]{Eisenhauer_others.2005}
{Eisenhauer} F.,  et~al., 2005, \apj, 628, 246

\bibitem[\protect\citeauthoryear{{Elias}, {Cabrera-Ca{\~n}o} \&
  {Alfaro}}{{Elias} et~al.}{2006}]{Elias_Cabrera-Cano.2006}
{Elias} F.,  {Cabrera-Ca{\~n}o} J.,    {Alfaro} E.~J.,  2006, \aj, 131, 2700

\bibitem[\protect\citeauthoryear{{Elmegreen} \& {Hunter}}{{Elmegreen} \&
  {Hunter}}{2006}]{Elmegreen_Hunter.2006}
{Elmegreen} B.~G.,  {Hunter} D.~A.,  2006, \apj, 636, 712

\bibitem[\protect\citeauthoryear{{Elmegreen} \& {Parravano}}{{Elmegreen} \&
  {Parravano}}{1994}]{Elmegreen_Parravano.1994}
{Elmegreen} B.~G.,  {Parravano} A.,  1994, \apjl, 435, L121+

\bibitem[\protect\citeauthoryear{{Erwin}, {Pohlen} \& {Beckman}}{{Erwin}
  et~al.}{2008}]{Erwin_Pohlen.2008}
{Erwin} P.,  {Pohlen} M.,    {Beckman} J.~E.,  2008, \aj, 135, 20

\bibitem[\protect\citeauthoryear{{Ferguson}, {Irwin}, {Chapman}, {Ibata},
  {Lewis} \& {Tanvir}}{{Ferguson} et~al.}{2007}]{Ferguson_Irwin.2007}
{Ferguson} A.,  {Irwin} M.,  {Chapman} S.,  {Ibata} R.,  {Lewis} G.,
  {Tanvir} N.,  2007, {Resolving the Stellar Outskirts of M31 and M33}.
pp 239--+

\bibitem[\protect\citeauthoryear{{Fitzpatrick}}{{Fitzpatrick}}{2004}]{Fitzpatr%
ick_only.2004}
{Fitzpatrick} E.~L.,  2004, in {Witt} A.~N.,  {Clayton} G.~C.,   {Draine}
  B.~T.,  eds, Astrophysics of Dust Vol.~309 of Astronomical Society of the
  Pacific Conference Series, {Interstellar Extinction in the Milky Way Galaxy}.
p.~33

\bibitem[\protect\citeauthoryear{{Freeman}}{{Freeman}}{1970}]{Freeman_only.197%
0}
{Freeman} K.~C.,  1970, \apj, 160, 811

\bibitem[\protect\citeauthoryear{{Freudenreich}}{{Freudenreich}}{1996}]{Freude%
nreich_only.1996}
{Freudenreich} H.~T.,  1996, \apj, 468, 663

\bibitem[\protect\citeauthoryear{{Friel}, {Janes}, {Tavarez}, {Scott},
  {Katsanis}, {Lotz}, {Hong} \& {Miller}}{{Friel}
  et~al.}{2002}]{Friel_Janes.2002}
{Friel} E.~D.,  {Janes} K.~A.,  {Tavarez} M.,  {Scott} J.,  {Katsanis} R.,
  {Lotz} J.,  {Hong} L.,    {Miller} N.,  2002, \aj, 124, 2693

\bibitem[\protect\citeauthoryear{{Ghez}, {Salim}, {Weinberg}, {Lu}, {Do},
  {Dunn}, {Matthews}, {Morris}, {Yelda}, {Becklin}, {Kremenek}, {Milosavljevic}
  \& {Naiman}}{{Ghez} et~al.}{2008}]{Ghez_Salim.2008}
{Ghez} A.~M.,  {Salim} S.,  {Weinberg} N.~N.,  {Lu} J.~R.,  {Do} T.,  {Dunn}
  J.~K.,  {Matthews} K.,  {Morris} M.~R.,  {Yelda} S.,  {Becklin} E.~E.,
  {Kremenek} T.,  {Milosavljevic} M.,    {Naiman} J.,  2008, \apj, 689, 1044

\bibitem[\protect\citeauthoryear{{Gillessen}, {Eisenhauer}, {Trippe},
  {Alexander}, {Genzel}, {Martins} \& {Ott}}{{Gillessen}
  et~al.}{2009}]{Gillessen_Eisenhauer.2009}
{Gillessen} S.,  {Eisenhauer} F.,  {Trippe} S.,  {Alexander} T.,  {Genzel} R.,
  {Martins} F.,    {Ott} T.,  2009, \apj, 692, 1075

\bibitem[\protect\citeauthoryear{{Gonz{\'a}lez-Solares}
  et~al.,}{{Gonz{\'a}lez-Solares}  et~al.}{2008}]{Gonzalez-Solares_others.2008}
{Gonz{\'a}lez-Solares} E.~A.,  et~al., 2008, \mnras, 388, 89

\bibitem[\protect\citeauthoryear{{Groot} et~al.,}{{Groot}
  et~al.}{2009}]{Groot_others.2009}
{Groot} P.,  et~al., 2009, ArXiv e-prints

\bibitem[\protect\citeauthoryear{{Habing}}{{Habing}}{1988}]{Habing_only.1988}
{Habing} H.~J.,  1988, \aap, 200, 40

\bibitem[\protect\citeauthoryear{{Hales}, {Barlow}, {Drew}, {Unruh}, {Greimel},
  {Irwin} \& {Gonz{\'a}lez-Solares}}{{Hales} et~al.}{2009}]{Hales_Barlow.2009}
{Hales} A.~S.,  {Barlow} M.~J.,  {Drew} J.~E.,  {Unruh} Y.~C.,  {Greimel} R.,
  {Irwin} M.~J.,    {Gonz{\'a}lez-Solares} E.,  2009, \apj, 695, 75

\bibitem[\protect\citeauthoryear{{Haywood}}{{Haywood}}{2006}]{Haywood_only.200%
6}
{Haywood} M.,  2006, \mnras, 371, 1760

\bibitem[\protect\citeauthoryear{{Holmberg}, {Nordstr{\"o}m} \&
  {Andersen}}{{Holmberg} et~al.}{2009}]{Holmberg_Nordstr"om.2009}
{Holmberg} J.,  {Nordstr{\"o}m} B.,    {Andersen} J.,  2009, \aap, 501, 941

\bibitem[\protect\citeauthoryear{{Houk}, {Swift}, {Murray}, {Penston} \&
  {Binney}}{{Houk} et~al.}{1997}]{Houk_Swift.1997}
{Houk} N.,  {Swift} C.~M.,  {Murray} C.~A.,  {Penston} M.~J.,    {Binney}
  J.~J.,  1997, in Hipparcos - Venice '97 Vol.~402 of ESA Special Publication,
  {The Properties of Main-Sequence Stars from HIPPARCOS Data}.
pp 279--282

\bibitem[\protect\citeauthoryear{{Joshi}}{{Joshi}}{2007}]{Joshi_only.2007}
{Joshi} Y.~C.,  2007, \mnras, 378, 768

\bibitem[\protect\citeauthoryear{{Juri{\'c}} et~al.,}{{Juri{\'c}}
  et~al.}{2008}]{Juric_others.2008}
{Juri{\'c}} M.,  et~al., 2008, \apj, 673, 864

\bibitem[\protect\citeauthoryear{{Kalberla}, {Burton}, {Hartmann}, {Arnal},
  {Bajaja}, {Morras} \& {P{\"o}ppel}}{{Kalberla}
  et~al.}{2005}]{Kalberla_Burton.2005}
{Kalberla} P.~M.~W.,  {Burton} W.~B.,  {Hartmann} D.,  {Arnal} E.~M.,  {Bajaja}
  E.,  {Morras} R.,    {P{\"o}ppel} W.~G.~L.,  2005, \aap, 440, 775

\bibitem[\protect\citeauthoryear{{Kennicutt}
  Jr.}{{Kennicutt}}{1989}]{Kennicutt_only.1989}
{Kennicutt} Jr. R.~C.,  1989, \apj, 344, 685

\bibitem[\protect\citeauthoryear{{Kent}, {Dame} \& {Fazio}}{{Kent}
  et~al.}{1991}]{Kent_Dame.1991}
{Kent} S.~M.,  {Dame} T.~M.,    {Fazio} G.,  1991, \apj, 378, 131

\bibitem[\protect\citeauthoryear{{Kregel}, {van der Kruit} \& {de
  Grijs}}{{Kregel} et~al.}{2002}]{Kregel_vanderKruit.2002}
{Kregel} M.,  {van der Kruit} P.~C.,    {de Grijs} R.,  2002, \mnras, 334, 646

\bibitem[\protect\citeauthoryear{{Larson}}{{Larson}}{1976}]{Larson_only.1976}
{Larson} R.~B.,  1976, \mnras, 176, 31

\bibitem[\protect\citeauthoryear{{Lemasle} et~al.,}{{Lemasle}
  et~al.}{2008}]{Lemasle_others.2008}
{Lemasle} B.,  et~al., 2008, Memorie della Societa Astronomica Italiana, 79,
  534

\bibitem[\protect\citeauthoryear{{Lemasle}, {Piersimoni}, {Pedicelli}, {Bono},
  {Fran{\c c}ois}, {Primas} \& {Romaniello}}{{Lemasle}
  et~al.}{2008}]{Lemasle_Piersimoni.2008}
{Lemasle} B.,  {Piersimoni} A.,  {Pedicelli} S.,  {Bono} G.,  {Fran{\c c}ois}
  P.,  {Primas} F.,    {Romaniello} M.,  2008, Memorie della Societa
  Astronomica Italiana, 79, 534

\bibitem[\protect\citeauthoryear{{Lewis} \& {Freeman}}{{Lewis} \&
  {Freeman}}{1989}]{Lewis_Freeman.1989}
{Lewis} J.~R.,  {Freeman} K.~C.,  1989, \aj, 97, 139

\bibitem[\protect\citeauthoryear{{Lucas} et~al.,}{{Lucas}
  et~al.}{2008}]{Lucas_others.2008}
{Lucas} P.~W.,  et~al., 2008, \mnras, 391, 136

\bibitem[\protect\citeauthoryear{{Luck}, {Kovtyukh} \& {Andrievsky}}{{Luck}
  et~al.}{2006}]{Luck_Kovtyukh.2006}
{Luck} R.~E.,  {Kovtyukh} V.~V.,    {Andrievsky} S.~M.,  2006, \aj, 132, 902

\bibitem[\protect\citeauthoryear{{Maciel} \& {Costa}}{{Maciel} \&
  {Costa}}{2009}]{Maciel_Costa.2009}
{Maciel} W.~J.,  {Costa} R.~D.~D.,  2009, in {Andersen} J.,  {Bland-Hawthorn}
  J.,   {Nordstr{\"o}m} B.,  eds, IAU Symposium Vol.~254 of IAU Symposium,
  {Abundance gradients in the galactic disk: Space and time variations}.
pp 38P--+

\bibitem[\protect\citeauthoryear{{Magrini}, {Corbelli} \& {Galli}}{{Magrini}
  et~al.}{2007}]{Magrini_Corbelli.2007}
{Magrini} L.,  {Corbelli} E.,    {Galli} D.,  2007, \aap, 470, 843

\bibitem[\protect\citeauthoryear{{Magrini}, {Perinotto}, {Mampaso} \&
  {Corradi}}{{Magrini} et~al.}{2004}]{Magrini_Perinotto.2004}
{Magrini} L.,  {Perinotto} M.,  {Mampaso} A.,    {Corradi} R.~L.~M.,  2004,
  \aap, 426, 779

\bibitem[\protect\citeauthoryear{{Magrini}, {Sestito}, {Randich} \&
  {Galli}}{{Magrini} et~al.}{2009}]{Magrini_Sestito.2009}
{Magrini} L.,  {Sestito} P.,  {Randich} S.,    {Galli} D.,  2009, \aap, 494, 95

\bibitem[\protect\citeauthoryear{{Magrini}, {Stanghellini} \&
  {Villaver}}{{Magrini} et~al.}{2009}]{Magrini_Stanghellini.2009}
{Magrini} L.,  {Stanghellini} L.,    {Villaver} E.,  2009, \apj, 696, 729

\bibitem[\protect\citeauthoryear{{Magrini}, {V{\'{\i}}lchez}, {Mampaso},
  {Corradi} \& {Leisy}}{{Magrini} et~al.}{2007}]{Magrini_Vilchez.2007}
{Magrini} L.,  {V{\'{\i}}lchez} J.~M.,  {Mampaso} A.,  {Corradi} R.~L.~M.,
  {Leisy} P.,  2007, \aap, 470, 865

\bibitem[\protect\citeauthoryear{{Marshall}, {Robin}, {Reyl{\'e}}, {Schultheis}
  \& {Picaud}}{{Marshall} et~al.}{2006}]{Marshall_Robin.2006}
{Marshall} D.~J.,  {Robin} A.~C.,  {Reyl{\'e}} C.,  {Schultheis} M.,
  {Picaud} S.,  2006, \aap, 453, 635

\bibitem[\protect\citeauthoryear{{Mart{\'{\i}}nez-Serrano}, {Serna},
  {Dom{\'e}nech-Moral} \&
  {Dom{\'{\i}}nguez-Tenreiro}}{{Mart{\'{\i}}nez-Serrano}
  et~al.}{2009}]{Martinez-Serrano_Serna.2009}
{Mart{\'{\i}}nez-Serrano} F.~J.,  {Serna} A.,  {Dom{\'e}nech-Moral} M.,
  {Dom{\'{\i}}nguez-Tenreiro} R.,  2009, ArXiv e-prints

\bibitem[\protect\citeauthoryear{{Matteucci} \& {Francois}}{{Matteucci} \&
  {Francois}}{1989}]{Matteucci_Francois.1989}
{Matteucci} F.,  {Francois} P.,  1989, \mnras, 239, 885

\bibitem[\protect\citeauthoryear{{Munari}, {Sordo}, {Castelli} \&
  {Zwitter}}{{Munari} et~al.}{2005}]{Munari_Sordo.2005}
{Munari} U.,  {Sordo} R.,  {Castelli} F.,    {Zwitter} T.,  2005, \aap, 442,
  1127

\bibitem[\protect\citeauthoryear{{Naab} \& {Ostriker}}{{Naab} \&
  {Ostriker}}{2006}]{Naab_Ostriker.2006}
{Naab} T.,  {Ostriker} J.~P.,  2006, \mnras, 366, 899

\bibitem[\protect\citeauthoryear{{Pflamm-Altenburg} \&
  {Kroupa}}{{Pflamm-Altenburg} \&
  {Kroupa}}{2008}]{Pflamm-Altenburg_Kroupa.2008}
{Pflamm-Altenburg} J.,  {Kroupa} P.,  2008, \nat, 455, 641

\bibitem[\protect\citeauthoryear{{Pietrinferni}, {Cassisi}, {Salaris} \&
  {Castelli}}{{Pietrinferni} et~al.}{2004}]{Pietrinferni_Cassisi.2004}
{Pietrinferni} A.,  {Cassisi} S.,  {Salaris} M.,    {Castelli} F.,  2004, \apj,
  612, 168

\bibitem[\protect\citeauthoryear{{Pohlen}, {L{\"u}tticke} \&
  {Dettmar}}{{Pohlen} et~al.}{2001}]{Pohlen_L"utticke.2001}
{Pohlen} M.,  {L{\"u}tticke} R.,    {Dettmar} R.-J.,  2001, in {Funes} J.~G.,
  {Corsini} E.~M.,  eds, Galaxy Disks and Disk Galaxies Vol.~230 of
  Astronomical Society of the Pacific Conference Series, {Cut-Off Radii of
  Galactic Disks}.
pp 135--136

\bibitem[\protect\citeauthoryear{{Pohlen} \& {Trujillo}}{{Pohlen} \&
  {Trujillo}}{2006}]{Pohlen_Trujillo.2006}
{Pohlen} M.,  {Trujillo} I.,  2006, \aap, 454, 759

\bibitem[\protect\citeauthoryear{{Reyl{\'e}}, {Marshall}, {Robin} \&
  {Schultheis{\'e}}}{{Reyl{\'e}} et~al.}{2009}]{Reyle_Marshall.2009}
{Reyl{\'e}} C.,  {Marshall} D.~J.,  {Robin} A.~C.,    {Schultheis{\'e}} M.,
  2009, \aap, 495, 819

\bibitem[\protect\citeauthoryear{{Robin}, {Creze} \& {Mohan}}{{Robin}
  et~al.}{1992}]{Robin_Creze.1992}
{Robin} A.~C.,  {Creze} M.,    {Mohan} V.,  1992, \aap, 265, 32

\bibitem[\protect\citeauthoryear{{Robin}, {Reyl{\'e}}, {Derri{\`e}re} \&
  {Picaud}}{{Robin} et~al.}{2003}]{Robin_Reyle.2003}
{Robin} A.~C.,  {Reyl{\'e}} C.,  {Derri{\`e}re} S.,    {Picaud} S.,  2003,
  \aap, 409, 523

\bibitem[\protect\citeauthoryear{{Ro{\v s}kar}, {Debattista}, {Stinson},
  {Quinn}, {Kaufmann} \& {Wadsley}}{{Ro{\v s}kar}
  et~al.}{2008}]{Rovskar_Debattista.2008}
{Ro{\v s}kar} R.,  {Debattista} V.~P.,  {Stinson} G.~S.,  {Quinn} T.~R.,
  {Kaufmann} T.,    {Wadsley} J.,  2008, \apjl, 675, L65

\bibitem[\protect\citeauthoryear{{Ruphy}, {Robin}, {Epchtein}, {Copet},
  {Bertin}, {Fouque} \& {Guglielmo}}{{Ruphy} et~al.}{1996}]{Ruphy_Robin.1996}
{Ruphy} S.,  {Robin} A.~C.,  {Epchtein} N.,  {Copet} E.,  {Bertin} E.,
  {Fouque} P.,    {Guglielmo} F.,  1996, \aap, 313, L21

\bibitem[\protect\citeauthoryear{{Sale} et~al.,}{{Sale}
  et~al.}{2009}]{Sale_others.2009}
{Sale} S.~E.,  et~al., 2009, \mnras, 392, 497

\bibitem[\protect\citeauthoryear{{Sanchez-Blazquez}, {Courty}, {Gibson} \&
  {Brook}}{{Sanchez-Blazquez} et~al.}{2009}]{Sanchez-Blazquez_Courty.2009}
{Sanchez-Blazquez} P.,  {Courty} S.,  {Gibson} B.,    {Brook} C.,  2009, ArXiv
  e-prints

\bibitem[\protect\citeauthoryear{{Santos}, {Melo}, {James}, {Gameiro},
  {Bouvier} \& {Gomes}}{{Santos} et~al.}{2008}]{Santos_Melo.2008}
{Santos} N.~C.,  {Melo} C.,  {James} D.~J.,  {Gameiro} J.~F.,  {Bouvier} J.,
  {Gomes} J.~I.,  2008, \aap, 480, 889

\bibitem[\protect\citeauthoryear{{Scalo}}{{Scalo}}{1986}]{Scalo_only.1986}
{Scalo} J.~M.,  1986, Fundamentals of Cosmic Physics, 11, 1

\bibitem[\protect\citeauthoryear{{Schaye}}{{Schaye}}{2004}]{Schaye_only.2004}
{Schaye} J.,  2004, \apj, 609, 667

\bibitem[\protect\citeauthoryear{{Searle}}{{Searle}}{1971}]{Searle_only.1971}
{Searle} L.,  1971, \apj, 168, 327

\bibitem[\protect\citeauthoryear{{Siegel}, {Majewski}, {Reid} \&
  {Thompson}}{{Siegel} et~al.}{2002}]{Siegel_Majewski.2002}
{Siegel} M.~H.,  {Majewski} S.~R.,  {Reid} I.~N.,    {Thompson} I.~B.,  2002,
  \apj, 578, 151

\bibitem[\protect\citeauthoryear{{Siess}, {Dufour} \& {Forestini}}{{Siess}
  et~al.}{2000}]{Siess_Dufour.2000}
{Siess} L.,  {Dufour} E.,    {Forestini} M.,  2000, \aap, 358, 593

\bibitem[\protect\citeauthoryear{{Trippe}, {Gillessen}, {Gerhard}, {Bartko},
  {Fritz}, {Maness}, {Eisenhauer}, {Martins}, {Ott}, {Dodds-Eden} \&
  {Genzel}}{{Trippe} et~al.}{2008}]{Trippe_Gillessen.2008}
{Trippe} S.,  {Gillessen} S.,  {Gerhard} O.~E.,  {Bartko} H.,  {Fritz} T.~K.,
  {Maness} H.~L.,  {Eisenhauer} F.,  {Martins} F.,  {Ott} T.,  {Dodds-Eden} K.,
     {Genzel} R.,  2008, \aap, 492, 419

\bibitem[\protect\citeauthoryear{{van der Kruit}}{{van der
  Kruit}}{1979}]{vanderKruit_only.1979}
{van der Kruit} P.~C.,  1979, \aaps, 38, 15

\bibitem[\protect\citeauthoryear{{van der Kruit}}{{van der
  Kruit}}{1986}]{vanderKruit_only.1986}
{van der Kruit} P.~C.,  1986, \aap, 157, 230

\bibitem[\protect\citeauthoryear{{van der Kruit}}{{van der
  Kruit}}{1987}]{vanderKruit_only.1987}
{van der Kruit} P.~C.,  1987, \aap, 173, 59

\bibitem[\protect\citeauthoryear{{Verley}, {Corbelli}, {Giovanardi} \&
  {Hunt}}{{Verley} et~al.}{2009}]{Verley_Corbelli.2009}
{Verley} S.,  {Corbelli} E.,  {Giovanardi} C.,    {Hunt} L.~K.,  2009, \aap,
  493, 453

\bibitem[\protect\citeauthoryear{{Viana Almeida}, {Santos}, {Melo}, {Ammler-von
  Eiff}, {Torres}, {Quast}, {Gameiro} \& {Sterzik}}{{Viana Almeida}
  et~al.}{2009}]{Viana_Almeida_Santos.2009}
{Viana Almeida} P.,  {Santos} N.~C.,  {Melo} C.,  {Ammler-von Eiff} M.,
  {Torres} C.~A.~O.,  {Quast} G.~R.,  {Gameiro} J.~F.,    {Sterzik} M.,  2009,
  \aap, 501, 965

\bibitem[\protect\citeauthoryear{{Vlaji{\'c}}, {Bland-Hawthorn} \&
  {Freeman}}{{Vlaji{\'c}} et~al.}{2009}]{Vlajic_Bland-Hawthorn.2009}
{Vlaji{\'c}} M.,  {Bland-Hawthorn} J.,    {Freeman} K.~C.,  2009, \apj, 697,
  361

\bibitem[\protect\citeauthoryear{{Williams}, {Dalcanton}, {Dolphin}, {Holtzman}
  \& {Sarajedini}}{{Williams} et~al.}{2009}]{Williams_Dalcanton.2009}
{Williams} B.~F.,  {Dalcanton} J.~J.,  {Dolphin} A.~E.,  {Holtzman} J.,
  {Sarajedini} A.,  2009, \apjl, 695, L15

\bibitem[\protect\citeauthoryear{{Worthey}, {Espa{\~n}a}, {MacArthur} \&
  {Courteau}}{{Worthey} et~al.}{2005}]{Worthey_Espana.2005}
{Worthey} G.,  {Espa{\~n}a} A.,  {MacArthur} L.~A.,    {Courteau} S.,  2005,
  \apj, 631, 820

\bibitem[\protect\citeauthoryear{{Yong}, {Carney} \& {Teixera de
  Almeida}}{{Yong} et~al.}{2005}]{Yong_Carney.2005}
{Yong} D.,  {Carney} B.~W.,    {Teixera de Almeida} M.~L.,  2005, \aj, 130, 597

\end{thebibliography}

\end{document}